\documentstyle[preprint,aps,eqsecnum]{revtex}

\preprint{ISSP April No. 2957, 1995, cond-mat/9504102}

\begin{document}

\title{Conductivity of 2D lattice electrons \\
in an incommensurate magnetic field}
\author{Masahito Takahashi, Yasuhiro Hatsugai$^\dagger$
\thanks{present address:
Department of Applied Physics, University of
Tokyo, 7-3-1 Hongo,  Bunkyo-ku, Tokyo 113, Japan.}
 and Mahito Kohmoto}
\date{March, 1995}
\address{Institute for Solid State Physics, University of Tokyo
7-22-1, Roppongi, Minato-ku, Tokyo 106, JAPAN}
\maketitle

\begin{abstract}
We consider conductivities of two-dimensional
lattice electrons in a magnetic field.
We focus on systems where the flux per plaquette
$\phi$ is irrational (incommensurate flux).
To realize the system with the incommensurate flux,
we consider a series of systems with commensurate fluxes which converge to
the irrational value. We have calculated a real part
of the longitudinal conductivity $\sigma_{xx}(\omega)$.
Using a scaling analysis, we have found $\Re\sigma_{xx}(\omega)$
behaves as $1/\omega ^{\gamma}$
\,$(\gamma =0.55)$ when $\phi =\tau,(\tau =\frac{\sqrt{5}-1}{2})$ and
the Fermi energy is near zero.
This behavior is closely related to the known
scaling behavior of the spectrum.
\end{abstract}


\section{Introduction}
\label{sec:intro}
The electronic properties of the two-dimensional periodic
systems in a
magnetic field have been studied extensively.
Even for non-interacting electrons,
the various physical quantities (for example, the
wavefunctions, and the energy spectra)
exhibit extremely rich behaviors
\cite{Harper,Wannier1,Azbel,Zak,Hofstadter,Wannier2,TKNN,Kohmoto2,Kohmoto3}
and it has been attracted great attentions
in relation to the
quantum Hall effect \cite{Prange,TKNN,Avron,Top,Rammal,Hatsugai0},
one-dimensional quasiperiodic systems
\cite{Aubry,Thouless,Kohmoto2,Ost,Kohmoto3},
flux states for the high-$T_c$
superconductivity \cite{WWZ,Affleck,Hasegawa,Hasegawa2}.
The algebraic structure of this problem has also been
revealed recently \cite{Wiegmann,Hatsugai1,Fadeev}.

Consider the tight-binding Hamiltonian on the square lattice
\begin{equation}
 H^{0}=-\sum_{<ij>}e^{i\theta_{ij}}c_{i}^{\dagger}c_{j} +{\rm h.c.},\\
        \label{eq:hamil}
\end{equation}
where $c_{i}^{\dagger} (c_{i})$ is
a creation (annihilation) operator of an electron
at site $i$.
The summation is taken over the nearest neighbor sites.
The hopping amplitude between link $\langle ij\rangle$
is set to be unity.
The phase factor $\theta_{ij}$ is defined on link
$\langle ij\rangle$.
The magnetic flux per plaquette is
$\phi=\frac{1}{2\pi}\sum_{\rm plaquette} \theta_{ij}$
in units of magnetic flux quantum $hc/e$.

When $\phi$ is rational, i.e. $\phi=p/q$ with mutual prime
integers $p$ and $q$, the spectrum consists
of $q$ bands with finite widths.
The wave functions are the extended Bloch
functions. Many interesting phenomena related to the quantum
Hall effects are discussed \cite{TKNN,Avron,Top,Rammal,Hatsugai0}.

When $\phi$ is irrational (incommensurate),
the system exhibits novel structures \cite{Hofstadter}.
It is known that the electronic state and the energy spectrum
have various singular natures.
The energy spectrum
is a Cartor set which
consists of infinitely many bands with zero width \cite{Aubry,Kohmoto2}.
Especially the system with $\phi=\tau\,(\tau =\frac{\sqrt{5}-1}{2})$
has been extensively studied \cite{Kohmoto2,Kohmoto3}.
The spectrum around $E=0$ shows a self-similar structure and
the clear scaling behavior is observed.
The wavefunctions
are critical and some of them show the multifractal
behavior \cite{Kohmoto3}.

In Ref. \cite{TKNN}, it is shown that
the Hall conductivity carried
is quantized to be an integer
in units of $e^2/h$ when the Fermi energy $E_F$ is in a gap.
This integer is given by the first Chern number
of the fiber bundle on the magnetic Brillouin zone \cite{Top}.
It is the total vorticity of the U(1) phase of the Bloch wavefunctions
which is a topological invariant.
On the other hand, the conductivity of the system
is also described by the edge states \cite{Rammal,Hatsugai0}.
It also has a topological
origin. The winding number of the edge states in the
complex energy plane gives the Hall conductivity \cite{Hatsugai0}.
Dou\c{c}ot and Stamp discussed
AC conductivities
for a commensurate flux $\phi=p/q$
when a particle density $\rho =\phi$ \cite{Stamp}.
In this case, the Fermi energy is in the largest gap
and the system is an insulator
\cite{Hasegawa,Hasegawa2}.

In this paper, we
consider both longitudinal and transverse
(Hall) conductivity. Especially we are interested in
the incommensurate flux limit.
When the flux $\phi$ is irrational, there are infinite
number of bands with zero width (Cantor set).
When the Fermi energy $E_F$ is
{\it not} in a gap with a finite width
in the Cantor set,
it is highly nontrivial whether the system is metallic or not.

In Sec.\ \ref{sec:cond}, we derive the expression for the
conductivity $\sigma_{\mu \nu,\phi}(\omega)$
for a commensuarate flux $\phi =p/q$
when $E_F$ is at an {\it arbitrary} position
in the spectrum using the Kubo formula.
In the following sections,
the longitudinal conductivity is discussed in details.
In Sec.\ \ref{sec:num}, systems with
a sequence of rational numbers $\{ \phi_l\}$ are
treated numerically in order
to understand the system in the incommensurate flux limit.
In Sec.\ \ref{sec:dis},
we discuss the incommensurate limit by taking an appropriate
scaling argument.
Sec.\ \ref{sec:sum} is a summary.
\section{Derivation of the conductivity}
\label{sec:cond}
In this section, we derive an expression for the conductivity
$\sigma_{\mu \nu ,\phi}(\omega)$
for a rational flux $\phi =p/q$
by the Kubo formula \cite{Kubo}. The real part
$\Re\sigma_{xx,\phi}(\omega)$ will be discussed in details.

Let us rewrite the Hamiltonian (\ref{eq:hamil}) as
\begin{eqnarray}
H^0&=&\sum_{\nu =x,y}H^0_{\nu}, \label {eq:hamil1} \\
H^0_{\nu}&=&-\sum_{\hat{l}} (e^{i\theta_{\hat{l}}^{\nu}}
c_{\hat{l}+\hat{\nu}}^{\dagger}c_{\hat{l}}+e^{-i\theta_{\hat{l}}^{\nu}}
c_{\hat{l}}^{\dagger}c_{\hat{l}+\hat{\nu}}). \label{eq:hamil2}
\end{eqnarray}
where
$\hat{l}$ denotes a two-dimensional lattice site
and $\hat{\nu}$ is a unit vector along the $\nu$ direction.
The phase factor
$\theta_{\hat{l}}^{\nu}$ is
on the link
between sites $\hat{l}$ and $\hat{l}+\hat{\nu}$.
We consider a response of the system to a uniform
time-dependent electric field $\mbox{\boldmath $E$}(t)
=(E^x(t),E^y(t))$.
Since the electric field $E^{\nu}_{ij}$
on a link $\langle ij \rangle$ is given by
$E^{\nu}_{ij}=-\frac{\hbar}{e} \frac{d\theta^{\nu}_{ij}}{dt}$,
the Hamiltonian $H$ with $\mbox{\boldmath $E$}(t)$
is given by changing the phase factor
$\theta_{\hat{l}}^{\nu}$ in Eq.~(\ref{eq:hamil2})
to $\theta_{\hat{l}}^{\nu}+\frac{e}{\hbar}A^{\nu}(t)$
with $-\frac{dA^{\nu}(t)}{dt}=E^{\nu}(t)$.
Up to the first order in $\mbox{\boldmath $E$}(t)$,
the Hamiltonian is expanded as
\begin{equation}
H=H^{0}+H'(t),\qquad H'(t)=-\sum_{\nu}J_{\nu}^{0}A^{\nu}(t),
\end{equation}
where
\begin{eqnarray}
J_{\mu}^{0}=i\frac{e}{\hbar}\sum_{\hat{l}} (e^{i\theta_{\hat{l}}^{\mu}}
c_{\hat{l}+\hat{\mu}}^{\dagger}c_{\hat{l}}-e^{-i\theta_{\hat{l}}^{\mu}}
c_{\hat{l}}^{\dagger}c_{\hat{l}+\hat{\mu}}),
\end{eqnarray}
which represents an unperturbed current operator along the $\mu$
direction.
The current operator is
\begin{eqnarray}
J_{\mu}=i\frac{e}{\hbar}\sum_{\hat{l}} (e^{i(\theta_{\hat{l}}^{\mu}
+\frac{e}{\hbar}A^{\mu}(t))}
 c_{\hat{l}+\hat{\mu}}^{\dagger}c_{\hat{l}}-
e^{-i(\theta_{\hat{l}}^{\mu}+\frac{e}{\hbar}A^{\mu}(t))}
c_{\hat{l}}^{\dagger}c_{\hat{l}+\hat{\mu}}).
\end{eqnarray}
It is expanded as
\begin{equation}
J_{\mu}=J_{\mu}^{0}+\frac{e^2}{\hbar ^2}H_{\mu}^{0}A^{\mu}(t),
\end{equation}
up to the first order in $\mbox{\boldmath $E$}(t)$.
By the Kubo formula, the induced electric current
$\delta \langle J_{\mu}(t)\rangle $ is
given by
\begin{eqnarray}
\delta \langle J_{\mu}(t)\rangle
=&&\frac{e^2}{\hbar ^2}
\langle H_{\mu}^{0}\rangle
A^{\mu}(t)+\frac{i}{\hbar}
\sum_{\nu}
\int_{0}^{+\infty}d\tau
\langle [J_{\mu}^{0}(\tau ), J_{\nu}^{0}]\rangle
A^{\nu}(t-\tau ), \\
J_{\mu}^{0}(\tau )=&&e^{i\frac{H^0 }{\hbar}\tau}J_{\mu}^0 e^{-i\frac{H^0
}{\hbar}\tau}.
\end{eqnarray}
where $\langle \cdots \rangle$ denotes the ground state expectation
value.
By Fourier transformation, we get
\begin{eqnarray}
\delta \langle J_{\mu}(\omega)\rangle
=\sum_{\nu}\bigg[2\pi i\frac{e^2}{h}\langle H_{\mu}^{0}\rangle\delta_{\mu \nu}
-K_{\mu \nu}(\omega )\bigg]\frac{1}{\hbar \omega -i0}E^{\nu}(\omega ),
\end{eqnarray}
with
\begin{equation}
K_{\mu \nu}(\omega)=\int_{0}^{+\infty }d\tau \langle
[J_{\mu}^{0}(\tau ), J_{\nu}^{0}]\rangle
e^{-i\omega \tau }. \label {eq:corr}
\end{equation}
Let us take the Landau gauge, that is,
$\theta_{\hat{l}}^{x}=0, \,\theta_{\hat{l}}^{y}=
2\pi \phi m$ where $\hat{l}=(m,n)$.
In a momentum representation,
we have
\begin{eqnarray}
 H^{0}&=&\int_{\rm M.B.Z.}
\frac{d^2k}{(2\pi )^{2}}\,
 \mbox{\boldmath $c$}^{\dagger}(\mbox{\boldmath $k$})
 \hat{\mbox{\boldmath $H$}}^{0}(\mbox{\boldmath $k$})
 \mbox{\boldmath $c$}(\mbox{\boldmath $k$}), \\
   J_{\mu}^{0}&=&\frac{e}{\hbar}\int_{\rm M.B.Z.}
\frac{d^2k}{(2\pi )^{2}}\,
 \mbox{\boldmath $c$}^{\dagger}(\mbox{\boldmath $k$})
 \hat{\mbox{\boldmath $v$}}_{\mu}(\mbox{\boldmath $k$})
 \mbox{\boldmath $c$}(\mbox{\boldmath $k$}), \label{eq:current}
\end{eqnarray}
with
\begin{eqnarray}
\hat{\mbox{\boldmath $H$}}^{0}(\mbox{\boldmath $k$})&=&
\sum_{\nu }\hat{\mbox{\boldmath $H$}}_{\nu}^{0}(\mbox{\boldmath $k$}),\\
\mbox{\boldmath $c$}^{\dagger}(\mbox{\boldmath $k$})&=&[c^{\dagger}(k_{x}+2\pi
\phi ,k_{y}), \cdots ,c^{\dagger}(k_{x}+2\pi \phi \,q,k_{y})], \\
 \{\hat{\mbox{\boldmath $H$}}_{x}^{0}(\mbox{\boldmath $k$})\}_{i,j}&=&
-2\cos (k_{x}+2\pi \phi j)\,\delta_{i,j}, \\
 \{\hat{\mbox{\boldmath $H$}}_{y}^{0}(\mbox{\boldmath $k$})\}_{i,j}&=&
 -(e^{-ik_y}\delta_{i,j+1}+e^{ik_y}\delta_{i,j-1}), \\
 \{\hat{\mbox{\boldmath $v$}}_{x}(\mbox{\boldmath $k$})\}_{i,j}&=&
  2\sin (k_{x}+2\pi \phi j)\,\delta_{i,j}, \\
 \{\hat{\mbox{\boldmath $v$}}_{y}(\mbox{\boldmath $k$})\}_{i,j}&=&
 i(e^{-ik_y}\delta_{i,j+1}-e^{ik_y}\delta_{i,j-1}),
\qquad(1 \le i, j \le q),
\end{eqnarray}
where $\mbox{\boldmath $k$}=(k_{x},k_{y})$
is defined in the magnetic Brillouin zone (M.B.Z.)
$(-\pi/q \le k_{x} < \pi/q, -\pi \le k_{y} < \pi )$,
and the operator
$c^{\dagger}(k_{x},k_{y})$ is defined by
$c^{\dagger}(k_{x},k_{y})=\sum_{m,n}e^{i(k_{x}m+k_{y}n)}c_{m,n}^{\dagger}$.

Let us discuss a one-particle state of the $l$-th band
\begin{equation}
|\Psi ^l\rangle =\int_{\rm M.B.Z.}\frac{d^2k}{(2\pi )^{2}}\,
\sum_{j=1}^{q}\psi_{j}^l(\mbox{\boldmath $k$})
c^{\dagger}(k_{x}+2\pi \phi j, k_{y})|0 \rangle.
\end{equation}
The Schr\"odinger equation
$H^{0}|\Psi ^l\rangle=E^l|\Psi ^l\rangle$
is reduced to
\begin{equation}
\hat{\mbox{\boldmath $H$}}^{0}(\mbox{\boldmath $k$})
|\mbox{\boldmath $k$}, l\rangle =E^{l}
(\mbox{\boldmath $k$})|\mbox{\boldmath $k$}, l\rangle ,
\end{equation}
with
\begin{equation}
|\mbox{\boldmath $k$}, l\rangle
=[\psi_{1}^l(\mbox{\boldmath $k$}), \cdots,
\psi_{q}^l(\mbox{\boldmath $k$})]^{\rm t}, \quad
\langle \mbox{\boldmath $k$}, l'|\mbox{\boldmath $k$}, l\rangle
=\delta_{l',l}.
\end{equation}

{}From Eqs.~(\ref{eq:current}) and (\ref{eq:corr}),
we have
\begin{eqnarray}
    K_{\mu \nu} (\omega )&=&V\frac{e^2}{h}\frac{i}{2\pi }
\int_{\rm M.B.Z.} d^2k\sum_{l,l'}
v_{\mu}^{ll'}(\mbox{\boldmath $k$})
v_{\nu}^{l'l}(\mbox{\boldmath $k$})
\frac{f^{l'}(\mbox{\boldmath $k$})-f^{l}(\mbox{\boldmath $k$})}
{\hbar \omega -i0+E^{l'}(\mbox{\boldmath $k$})
-E^{l}(\mbox{\boldmath $k$})}, \\
 v_{\mu}^{ll'}(\mbox{\boldmath $k$})
 &=&\langle \mbox{\boldmath $k$},l|
\hat{\mbox{\boldmath $v$}}_{\mu}(\mbox{\boldmath $k$})
|\mbox{\boldmath $k$}, l'\rangle .
\end{eqnarray}
Similarly we have
\begin{eqnarray}
\langle H_{\mu}^{0}\rangle &=&V
\int_{\rm M.B.Z.} \frac{d^2k}{(2\pi )^2}
\sum_{l} \epsilon_{\mu}^{l}(\mbox{\boldmath $k$})
f^{l}(\mbox{\boldmath $k$}), \\
\epsilon_{\mu}^{l}(\mbox{\boldmath $k$})&=&
\langle \mbox{\boldmath $k$},l|
 \hat{\mbox{\boldmath $H$}}_{\mu}^{0}(\mbox{\boldmath $k$})
|\mbox{\boldmath $k$}, l\rangle ,
\end{eqnarray}
where $V$ is the volume of the system and $f^{l}(\mbox{\boldmath $k$})
=\theta (E_{F}-E^{l}(\mbox{\boldmath $k$}))$.
Since $\delta \langle J_{\mu}(\omega )\rangle
=V \,\frac{e^2}{h}\, \sum_{\nu}\sigma_{\mu \nu,\phi}(\omega )
E^{\nu}(\omega )$,
we have
\begin{eqnarray}
    \sigma_{\mu \nu,\phi} (\omega )&=&-\frac{1}{2\pi i}\int_{\rm M.B.Z.}
d^2k\sum_{l}\epsilon_{\mu}^{l}(\mbox{\boldmath $k$})f^{l}(\mbox{\boldmath $k$})
\frac{1}{\hbar \omega -i0}\delta_{\mu \nu} \nonumber \\
&+&\frac{1}{2\pi i} \int_{\rm M.B.Z.} d^2k
\sum_{l,l'} v_{\mu}^{ll'}(\mbox{\boldmath $k$})
v_{\nu}^{l'l}(\mbox{\boldmath $k$})
\frac{f^{l'}(\mbox{\boldmath $k$})-f^{l}(\mbox{\boldmath $k$})}{\hbar \omega
-i0+E^{l'}(\mbox{\boldmath $k$})-E^{l}(\mbox{\boldmath $k$})}
\, \frac{1}{\hbar \omega -i0}. \label{eq:sigma1}
\end{eqnarray}
Let us rewrite the second term as
\begin{eqnarray}
\frac{1}{2\pi i} \int_{\rm M.B.Z.} d^2k
\sum_{l\ne l'} v_{\mu}^{ll'}(\mbox{\boldmath $k$})
v_{\nu}^{l'l}(\mbox{\boldmath $k$})
\frac{f^{l'}(\mbox{\boldmath $k$})-f^{l}(\mbox{\boldmath $k$})}
{E^{l'}(\mbox{\boldmath $k$})-E^{l}(\mbox{\boldmath $k$})}
\bigg(\frac{1}{\hbar \omega -i0}
-\frac{1}{\hbar \omega -i0+E^{l'}(\mbox{\boldmath $k$})-E^{l}(\mbox{\boldmath
$k$})}\bigg ) \nonumber \\
=
\frac{1}{2\pi i} \int_{\rm M.B.Z.} d^2k
\sum_{l}
f^{l}(\mbox{\boldmath $k$})\sum_{l' \ne l} \bigg[v_{\nu}^{ll'}(\mbox{\boldmath
$k$})
\frac{v_{\mu}^{l'l}(\mbox{\boldmath $k$})}
{E^{l}(\mbox{\boldmath $k$})-E^{l'}(\mbox{\boldmath $k$})}-
\frac{v_{\mu}^{ll'}(\mbox{\boldmath $k$})}
{E^{l'}(\mbox{\boldmath $k$})-E^{l}(\mbox{\boldmath $k$})}
v_{\nu}^{l'l}(\mbox{\boldmath $k$})\bigg]
\frac{1}{\hbar \omega -i0} \nonumber \\
+\frac{1}{2\pi i} \int_{\rm M.B.Z.} d^2k
\sum_{l\ne l'} v_{\mu}^{ll'}(\mbox{\boldmath $k$})
v_{\nu}^{l'l}(\mbox{\boldmath $k$})
\frac{f^{l'}(\mbox{\boldmath $k$})-f^{l}(\mbox{\boldmath $k$})}
{E^{l}(\mbox{\boldmath $k$})-E^{l'}(\mbox{\boldmath $k$})}
\frac{1}{\hbar \omega -i0+E^{l'}(\mbox{\boldmath $k$})-E^{l}(\mbox{\boldmath
$k$})}. \label{eq:chan}
\end{eqnarray}
Using the formulae
\begin{equation}
\langle \mbox{\boldmath $k$},l|
\frac{\partial }{\partial k_{\mu}}
|\mbox{\boldmath $k$}, l' \rangle =
 \frac{v_{\mu}^{ll'}(\mbox{\boldmath $k$})}
{E^{l'}(\mbox{\boldmath $k$})-E^{l}(\mbox{\boldmath $k$})}\quad (l \ne l'),
\label{eq:formula1}
\end{equation}
and
\begin{equation}
 v_{\mu}^{ll}(\mbox{\boldmath $k$})=
\frac{\partial E^{l}(\mbox{\boldmath $k$})}{\partial k_\mu},
\end{equation}
the first term of Eq.~(\ref{eq:chan}) is written
\begin{eqnarray}
\frac{1}{2\pi i} \int_{\rm M.B.Z.} d^2k&&
\sum_{l}
f^{l}(\mbox{\boldmath $k$})
\sum_{l'\ne l}\bigg[ v_{\nu}^{ll'}(\mbox{\boldmath $k$})
\langle \mbox{\boldmath $k$},l'|
\frac{\partial }{\partial k_{\mu}}
|\mbox{\boldmath $k$},l\rangle
-\langle \mbox{\boldmath $k$},l|
\frac{\partial }{\partial k_{\mu}}
|\mbox{\boldmath $k$},l'\rangle
v_{\nu}^{l'l}(\mbox{\boldmath $k$})
\bigg]\frac{1}{\hbar \omega -i0} \nonumber \\
=\frac{1}{2\pi i} \int_{\rm M.B.Z.} d^2k&& \nonumber \\
\times \sum_{l}
f^{l}(\mbox{\boldmath $k$})
\bigg[
\langle \mbox{\boldmath $k$},l|
&&\hat{\mbox{\boldmath $v$}}_{\nu}(\mbox{\boldmath $k$})
\frac{\partial }{\partial k_{\mu}}
|\mbox{\boldmath $k$},l\rangle
-\langle \mbox{\boldmath $k$},l|
\frac{\partial }{\partial k_{\mu}}
\bigg( \hat{\mbox{\boldmath $v$}}_{\nu}(\mbox{\boldmath $k$})
|\mbox{\boldmath $k$},l \rangle \bigg)
+\frac{\partial }{\partial k_{\mu}}
\langle \mbox{\boldmath $k$},l|
\hat{\mbox{\boldmath $v$}}_{\nu}(\mbox{\boldmath $k$})
|\mbox{\boldmath $k$},l\rangle \bigg]
\frac{1}{\hbar \omega -i0} \nonumber \\
=\frac{1}{2\pi i} \int_{\rm M.B.Z.} d^2k
&&\sum_{l}
f^{l}(\mbox{\boldmath $k$})
\bigg[-\langle \mbox{\boldmath $k$},l|
\frac{\partial \hat{\mbox{\boldmath $v$}}_{\nu}(\mbox{\boldmath $k$})}
{\partial k_{\mu}}
|\mbox{\boldmath $k$},l\rangle
+\frac{\partial v_{\nu}^{ll}(\mbox{\boldmath $k$})}{\partial k_{\mu}}\bigg]
\frac{1}{\hbar \omega -i0} \nonumber \\
=\frac{1}{2\pi i} \int_{\rm M.B.Z.} d^2k
&&\sum_{l}
f^{l}(\mbox{\boldmath $k$})
\epsilon_{\mu}^{l}(\mbox{\boldmath $k$})\delta_{\mu \nu}
\frac{1}{\hbar \omega -i0}
-\frac{1}{2\pi i} \int_{\rm M.B.Z.} d^2k
\sum_{l}
{|v_{\mu}^{ll}(\mbox{\boldmath $k$})|^{2}}
\frac{\partial f^{l}(\mbox{\boldmath $k$})}{\partial E}
\delta_{\mu \nu}\frac{1}{\hbar \omega -i0}.
\label {eq:trans}
\end{eqnarray}
Next let us separate the $\omega$-dependent part
in the second term of Eq.~(\ref{eq:chan}) and we have
\begin{eqnarray}
&&\frac{1}{2\pi i} \int_{\rm M.B.Z.} d^2k
\sum_{l\ne l'} v_{\mu}^{ll'}(\mbox{\boldmath $k$})
v_{\nu}^{l'l}(\mbox{\boldmath $k$})
\frac{f^{l'}(\mbox{\boldmath $k$})-f^{l}(\mbox{\boldmath $k$})}
{E^{l}(\mbox{\boldmath $k$})-E^{l'}(\mbox{\boldmath $k$})}
\frac{1}{\hbar \omega -i0+E^{l'}(\mbox{\boldmath $k$})-
E^{l}(\mbox{\boldmath $k$})} \nonumber \\
=&&\frac{1}{2\pi i} \int_{\rm M.B.Z.} d^2k
\sum_{l\ne l'} v_{\mu}^{ll'}(\mbox{\boldmath $k$})
v_{\nu}^{l'l}(\mbox{\boldmath $k$})
\frac{f^{l}(\mbox{\boldmath $k$})-f^{l'}(\mbox{\boldmath $k$})}
{(E^{l}(\mbox{\boldmath $k$})
-E^{l'}(\mbox{\boldmath $k$}))^2}   \nonumber \\
&&+\frac{1}{2\pi i} \int_{\rm M.B.Z.} d^2k
\sum_{l\ne l'} v_{\mu}^{ll'}(\mbox{\boldmath $k$})
v_{\nu}^{l'l}(\mbox{\boldmath $k$})
\frac{f^{l'}(\mbox{\boldmath $k$})-f^{l}(\mbox{\boldmath $k$})}
{(E^{l}(\mbox{\boldmath $k$})-E^{l'}(\mbox{\boldmath $k$}))^2}
\frac{\hbar \omega}{\hbar \omega -i0+E^{l'}(\mbox{\boldmath $k$})
-E^{l}(\mbox{\boldmath $k$})}, \label{eq:trans1}
\end{eqnarray}
Using Eq.~(\ref{eq:formula1}) and
$\langle \mbox{\boldmath $k$},l|\frac{\partial }{\partial k_{\nu}}
|\mbox{\boldmath $k$},l'\rangle = -
\bigg(\frac{\partial }{\partial k_{\nu}}
\langle \mbox{\boldmath $k$},l|\bigg)
|\mbox{\boldmath $k$},l'\rangle$, the first term of
Eq.~(\ref{eq:trans1}) is written
\begin{eqnarray}
&&\frac{1}{2\pi i} \int_{\rm M.B.Z.} d^2k
\sum_{l}
f^{l}(\mbox{\boldmath $k$})
\sum_{l'\ne l}\bigg[
\frac{v_{\mu}^{ll'}(\mbox{\boldmath $k$})
v_{\nu}^{l'l}(\mbox{\boldmath $k$})
-v_{\nu}^{ll'}(\mbox{\boldmath $k$})
v_{\mu}^{l'l}(\mbox{\boldmath $k$})}
{(E^{l}(\mbox{\boldmath $k$})-E^{l'}(\mbox{\boldmath $k$}))^2}
\bigg] \nonumber \\
=&&\frac{1}{2\pi i} \int_{\rm M.B.Z.} d^2k
\sum_{l}
f^{l}(\mbox{\boldmath $k$})
\sum_{l'}\bigg[
\langle \mbox{\boldmath $k$},l|\frac{\partial }
{\partial k_{\nu}}|\mbox{\boldmath $k$},l'\rangle
\langle \mbox{\boldmath $k$},l'|
\frac{\partial }{\partial k_{\mu}}
|\mbox{\boldmath $k$},l\rangle
-\langle \mbox{\boldmath $k$},l|
\frac{\partial }{\partial k_{\mu}}
|\mbox{\boldmath $k$},l' \rangle
\langle \mbox{\boldmath $k$},l'|
\frac{\partial }{\partial k_{\nu}}
|\mbox{\boldmath $k$},l\rangle
\bigg] \nonumber \\
=&&\frac{1}{2\pi i} \int_{\rm M.B.Z.} d^2k
\sum_{l}
f^{l}(\mbox{\boldmath $k$})
\bigg[
\bigg(\frac{\partial }{\partial k_{\mu}}
\langle \mbox{\boldmath $k$},l|\bigg)
\frac{\partial }{\partial k_{\nu}}
|\mbox{\boldmath $k$},l\rangle
-\bigg(\frac{\partial }{\partial k_{\nu}}
\langle \mbox{\boldmath $k$},l|\bigg)
\frac{\partial }{\partial k_{\mu}}
|\mbox{\boldmath $k$},l\rangle
\bigg] \nonumber \\
=&&\frac{1}{2\pi i} \int_{\rm M.B.Z.} d^2k
\sum_{l}
f^{l}(\mbox{\boldmath $k$})
\,\nabla \times
\langle \mbox{\boldmath $k$},l|
\nabla
|\mbox{\boldmath $k$},l\rangle \,\epsilon_{\mu \nu}
\label{eq:tknn}
\end{eqnarray}
{}From Eqs.~(\ref{eq:sigma1}), (\ref{eq:trans}), (\ref{eq:trans1})
and (\ref{eq:tknn})
with ${\partial f^{l}(\mbox{\boldmath $k$})}/{\partial E}=
-\delta (E_F-E^{l}(\mbox{\boldmath $k$}))$, we have
\begin{eqnarray}
\sigma_{\mu \nu ,\phi}(\omega)=&&
\frac{1}{2\pi i} \int_{\rm M.B.Z.} d^2k
\sum_{l}
{|v_{\mu}^{ll}(\mbox{\boldmath $k$})|^{2}}
\delta (E_F-E^{l}(\mbox{\boldmath $k$}))
\frac{1}{\hbar \omega -i0}\delta_{\mu \nu} \nonumber \\
+&&\frac{1}{2\pi i} \int_{\rm M.B.Z.} d^2k \sum_{l}
f^{l}(\mbox{\boldmath $k$})
\,\nabla \times
\langle \mbox{\boldmath $k$},l|
\nabla
|\mbox{\boldmath $k$},l\rangle \,\epsilon_{\mu \nu} \nonumber \\
+&&\frac{1}{2\pi i} \int_{\rm M.B.Z.} d^2k
\sum_{l\ne l'} v_{\mu}^{ll'}(\mbox{\boldmath $k$})
v_{\nu}^{l'l}(\mbox{\boldmath $k$})
\frac{f^{l'}(\mbox{\boldmath $k$})-f^{l}(\mbox{\boldmath $k$})}
{(E^{l}(\mbox{\boldmath $k$})-E^{l'}(\mbox{\boldmath $k$}))^2}
\frac{\hbar \omega}{\hbar \omega -i0+E^{l'}(\mbox{\boldmath $k$})
-E^{l}(\mbox{\boldmath $k$})} \label{eq:sigma4} .
\end{eqnarray}
For the Hall conductivity $\sigma_{xy,\phi} (\omega )$,
we have
\begin{eqnarray}
\sigma_{xy,\phi} (\omega ) &=& \sigma_{xy,\phi}^{{\rm TKNN}}+
\sigma_{xy,\phi}^{S} (\omega ), \\
\sigma_{xy,\phi}^{{\rm TKNN}}\,\,\,\,
&=&\frac{1}{2\pi i} \int_{\rm M.B.Z.} d^2k \sum_{l}
f^{l}(\mbox{\boldmath $k$})
\,\nabla \times
\langle \mbox{\boldmath $k$},l|
\nabla
|\mbox{\boldmath $k$},l\rangle , \\
\sigma_{xy,\phi}^{S} (\omega ) &=&\frac{1}{2\pi i} \int_{\rm M.B.Z.} d^2k
\sum_{l\ne l'} v_{x}^{ll'}(\mbox{\boldmath $k$})
v_{y}^{l'l}(\mbox{\boldmath $k$})
\frac{f^{l'}(\mbox{\boldmath $k$})-f^{l}(\mbox{\boldmath $k$})}
{(E^{l}(\mbox{\boldmath $k$})-E^{l'}(\mbox{\boldmath $k$}))^2}
\frac{\hbar \omega}{\hbar \omega -i0+E^{l'}(\mbox{\boldmath $k$})
-E^{l}(\mbox{\boldmath $k$})}.
\end{eqnarray}
Since
$\sigma_{xy,\phi}^{S} (\omega =0)=0$, the static Hall
conductivity is given by $\sigma_{xy,\phi}^{{\rm TKNN}}$ which
is a topological invariant in the M.B.Z.

For the longitudinal conductivity $\sigma_{xx,\phi} (\omega )$,
we have
\begin{eqnarray}
    \sigma_{xx,\phi} (\omega )&=&\sigma_{xx,\phi}^D(\omega)
+\sigma_{xx,\phi}^S(\omega), \\
\sigma_{xx,\phi}^D(\omega)
&=&\frac{1}{2\pi i}\int_{\rm M.B.Z.}
d^2k
\sum_{l}{| v_{x}^{ll}(\mbox{\boldmath $k$})|}^{2}
\delta (E_F-E^{l}(\mbox{\boldmath $k$}))
\frac{1}{\hbar \omega -i0}, \\
\sigma_{xx,\phi }^S(\omega )&=
&\frac{1}{2\pi i} \int_{\rm M.B.Z.} d^2k
\sum_{l\ne l'} {\mid v_{x}^{ll'}(\mbox{\boldmath $k$})\mid}^{2}
\frac{f^{l'}(\mbox{\boldmath $k$})-f^{l}(\mbox{\boldmath $k$})}
{(E^{l}(\mbox{\boldmath $k$})-E^{l'}(\mbox{\boldmath $k$}))^2}
\frac{\hbar \omega}{\hbar \omega -i0+E^{l'}(\mbox{\boldmath
$k$})-E^{l}(\mbox{\boldmath $k$})},
\end{eqnarray}
where $\sigma_{xx,\phi}^D(\omega)$ is the so-called Drude term
and $\sigma_{xx,\phi}^S(\omega)$
is the contribution from interband scattering processes.
Real parts of them are evaluated as
\begin{eqnarray}
    \Re\sigma_{xx,\phi}^D(\omega )=&&D(q)\delta (\hbar \omega), \\
D(q)=&&\frac{1}{2} \int_{\rm M.B.Z.} d^2k
\sum_{l} {| v_{x}^{ll}(\mbox{\boldmath $k$})|}^{2}
\delta (E_F-E^{l}(\mbox{\boldmath $k$})),  \nonumber \\
  =&&\left\{
\begin{array}{@{\,}ll}
\frac{1}{2} \oint_{E^{m}(k)=E_F}dk
{| v_{x}^{mm}(\mbox{\boldmath $k$})|}^{2}
\frac{1}{|\nabla E^{m}(k)|}
& \mbox{($E_F$ is in the $m$-th band)} \\
0 & \mbox{($E_F$ is in the energy gap)},
  \end{array}
  \right.    \label{eq:Drude}
\end{eqnarray}
and
\begin{eqnarray}
\Re\sigma_{xx,\phi}^S(\omega )=
 &&\frac{1}{2} \int_{\rm M.B.Z.} d^2k\sum_{l \ne l'} {\mid
v_{x}^{ll'}(\mbox{\boldmath $k$})\mid}^{2}\{f^{l'}(\mbox{\boldmath
$k$})-f^{l}(\mbox{\boldmath $k$})\}
\delta (\hbar \omega +E^{l'}(\mbox{\boldmath $k$})-E^{l}(\mbox{\boldmath $k$}))
\frac{1}{\hbar \omega }, \nonumber \\
=&&\frac{1}{2\hbar \omega}\sum_{l \ne l',E^l \ge E_F \ge E^{l'}}
\oint_{E^l(k) -E^{l'} (k) =\hbar \omega } dk
{|v_{x}^{ll'}(\mbox{\boldmath $k$})|}^{2}
\frac{1}{|\nabla (E^{l}(k)-E^{l'}(k))|}.
\label{eq:scatter}
\end{eqnarray}
The Drude term $\Re\sigma^D_{xx,\phi}(\omega )$
is a delta function at $\omega =0$
with the weight
being determined by the average velocity of the electrons
at the Fermi lines (there are $q$ Fermi lines).
The second term $\Re\sigma^S_{xx,\phi}(\omega )$ comes from
the interband scattering processes between states
$(\mbox{\boldmath $k$},l')$ and $(\mbox{\boldmath $k$},l)$.

For a rational flux, the onset of $\Re\sigma^S_{xx,\phi}(\omega )$ is
given by the minimum energy gap.
When the Fermi energy is in the energy band,
the Drude term $\Re\sigma^D_{xx,\phi}(\omega )$ can also
contribute.
\section{Numerical results for the irrational flux}
\label{sec:num}
In order to understand the incommensurate flux case,
we approximate the irrational $\phi$ by a series of rational fluxes
which tends to the irrational as $q \to \infty$.
For a large $q$, the onset of the $\Re\sigma^S_{xx,\phi}(\omega )$
becomes small and it is difficult to
distinguish $\Re\sigma^S_{xx,\phi}(\omega )$
from $\Re\sigma^D_{xx,\phi}(\omega )$.
Thus careful considerations are
necessary to understand the behavior of
$\Re\sigma_{xx,\phi}(\omega )$.

Let us consider the
systems with $\phi_l =F_l/F_{l+1}$
where $F_l$ is the Fibonacci number
defined by $F_{l+1}=F_{l}+F_{l-1}$ with $F_{0}=F_{1}=1$.
The flux $\phi_l$ tends to $\phi =\frac{\sqrt{5}-1}{2}$
(the inverse of the golden mean for $l \to \infty$).
Calculations were done for $l=5 \sim 12$ which correspond to the fluxes
$\frac{8}{13}, \frac{13}{21}, \frac{21}{34}, \frac{34}{55},
\frac{55}{89}, \frac{89}{144},
\frac{144}{233}$ and $\frac{233}{377}$.
The one-dimensional integrations in Eq.~(\ref{eq:Drude}) and
Eq.~(\ref{eq:scatter}) were
numerically performed by the double exponential formula.

First $E_F$ was put in an energy band.
We kept $E_F$ near zero since there is a energy band around
$E=0$ if $q$ is odd.
The cases with even $q$ were avoided since
the density of states disappear linearly at $E=0$
\cite{Wen,Kohmoto1,Hatsugai2}.
The Drude weights $D(q)$ are shown in Fig.1.
We treated three systems with
$E_F=0$, $E_F=-0.475\Delta_l$
and $E_F=-0.4995\Delta_l$
where $\Delta_l$ is a width
of the central band in the $l$-th stage.
In some cases, we have data for $\phi =610/987(l=14)$.
Since it is known that the spectrum shows a self-similar
structure for every three fluxes $(\phi_l,
\phi_{l+3}, \cdots )$ \cite{Kohmoto2},
we fitted the results within these
groups.
Numerical results are well fitted by the following scaling form
\begin{equation}
        D(q_l) = \bar{D}\,\frac{1}{q_l^{\alpha}},
       \qquad\alpha =0.828 \pm 0.002.
  \label{eq:alpha}
\end{equation}
where the constant $\bar{D}$ depends on the position of
$E_F$ but $\alpha$ is universal (see Fig.1).
We calculated the interband scattering term
$\Re\sigma_{xx,\phi_l}^S(\omega )$ when $E_F=0$
and they are shown in Figs.2.
In the $q_l \to \infty$ limit, the energy gap near $E=0$
becomes infinitesimally small and the interband scattering term
$\Re\sigma_{xx,\phi_l}^S(\omega )$
can also contribute in $\omega \sim 0$.
Next let us put $E_F$ in the energy gap near $E=0$.
We have calculated
$\Re\sigma_{xx,\phi_l}^S(\omega )$
when
$E_F$ is in the gap just above $E=0$.
Numerical calculations for even $q$ are also included.
In this case, the Drude term
does not contribute and we have calculated only
$\Re\sigma^{S}_{xx,\phi_l}(\omega )$. Numerical results
are shown in Figs.3.

As shown in Figs.2 and Figs.3,
the self-similar structure
is reflected on the $\omega$-dependence
of $\Re\sigma_{xx,\phi_l}^S(\omega )$.
This is consistent with the
the self-similar behavior observed in the energy spectrum.
The supports of
$\Re\sigma_{xx,\phi_l}^S(\omega )$ vanish when
the flux approaches to the irrational value.
\section{discussions}
\label{sec:dis}
Since $\Re\sigma_{xx,\phi_l}(\omega )$ shows the complex
behavior,
we analyze the numerical results in Sec.\ \ref{sec:num}
carefully and discuss scaling behaviors
of the longitudinal conductivity.
Since $\phi_l$ can be considered as $l$-th approximation of
the irrational flux $\phi=\tau$,
the energy resolution of the calculations
should be determined
by the spectrum in the $l$-th approximation. Thus
$\Re\sigma_{xx,\phi_l}^{(D,S)}(\omega )$
is averaged over the energy window $[\omega -W_l/2, \omega +W_l/2]$
to define
$\Re\sigma_{{\rm ave},\phi_l}^{(D,S)}(\omega )$
\begin{eqnarray}
        \Re\sigma_{{\rm ave},\phi_l}^{(D, S)}(\omega)=
        \frac{1}{W_l}\int_{\omega -W_l/2}^{\omega +W_l/2}
          \Re\sigma_{xx,\phi_l}^{(D, S)}(\omega ' )\,d\omega'\, ,
\label{eq:ave1}
\end{eqnarray}
where $W_l$ is determined by the energy resolution of the energy spectrum.
We put $\omega =0$ to consider
the small $\omega $ behavior.
Denote $\Re\sigma_{{\rm ave},\phi_l}^{(D, S)}(0)$ by
$\Re\sigma_{\rm ave}^{(D, S)}(\phi_l)$.
We take $W_l$
the same as the energy scale $\Delta E_l$ of the energy spectrum around $E=0$.
Let us take $\Delta E_l$ to be the energy gap around $E=0$.
The scaling form of $\Delta E_l$ is numerically given by
\begin{equation}
         \Delta E_l=\frac{{\rm const.}}{q_l^{\eta}},
         \qquad\eta=1.8285 \pm 0.0006. \label{eq:del}
\end{equation}
This scaling behavior
was already discussed in Ref. \cite{Kohmoto2}.
We choose the window $W_l$ as
\begin{equation}
         W_l=\frac{W_{0}}{q_l^{\eta}}, \label{eq:window}
\end{equation}
where $W_0$ is a $q$-independent constant.
The number of bands $M$ within the window $W_l$ is determined by $W_0$
which is independent of $l$.
The average conductivities
$\Re\sigma_{\rm ave}(\phi_l)$ were calculated for
three values of $W_0$
which give
$M=3, 7$, and $11$ respectively.
We consider the two cases separately.
One is when the Fermi energy is in the energy gap
and the other is when the Fermi energy is in the energy band.
First let us consider the former case.
The averaged values of $\Re\sigma^S_{\rm ave}(\phi_l)$
are shown in Fig.4.(a).
The results suggest strongly the
scaling behavior
\begin{equation}
\Re\sigma^S_{\rm ave}(\phi_l)=c(W_0)\,q_l^{\delta_S},
\qquad \delta_S=0.99 \pm 0.02. \label{eq:scale1}
\end{equation}
The coefficient $c(W_0)$ depends on the system
but it is independent of $q_l$.
Next let us consider the latter case when $E_F$ is in the band.
We calculated $\Re\sigma_{\rm ave}^{S}(\phi_l)$
similarly.
The results
are shown in Fig.4.(b).
They also suggest the scaling form Eq.~(\ref{eq:scale1})
with the exponent $\delta_S =0.99 \pm 0.07$.
In this case, the Drude term
$\Re\sigma_{{xx},\phi_l}^{D}(\omega )=
D(q)\delta (\hbar \omega )$ also
contributes to $\Re\sigma_{\rm ave}(\phi_l)$
since the Fermi energy is in the energy band.
The contribution from the Drude term is obtained as
\begin{equation}
        \Re\sigma^D_{\rm ave}(\phi_l)=
        \frac{1}{W_l}\int_{-W_l/2}^{W_l/2}
          \Re\sigma^D_{xx,\phi_l}(\omega ')\,d\omega '
                   =c'(W_0)\,q_l^{\delta_D}, \label{eq:deltad}
\end{equation}
where $c'(W_0)$ is independent of $q_l$.
The scaling exponent $\alpha$ for the Drude weight is
defined from Eqs.~(\ref{eq:window}) and (\ref{eq:deltad}) to be
\begin{equation}
\alpha= \eta -\delta_D. \label{eq:Drude00}
\end{equation}
The value of the exponent $\delta_D$ is identical to
$\delta_S$ within our numerical accuracy.

Define $\delta \equiv \delta_S \simeq \delta_D$, then
the average
conductivity $\Re\sigma_{\rm ave}(\phi_l)$ has
the scaling form
\begin{equation}
\Re\sigma_{\rm ave}(\phi_l)=c(W_0)\,q_l^{\delta}.
\label{eq:scale00}
\end{equation}
This form is universal in the sense that the exponent $\delta$
is independent of the position of $E_F$ and the way the
approximation of the irrational flux is made. Furthermore, it does not
depend on whether the Fermi energy is in an energy gap or not.
{}From Eqs.~(\ref{eq:del}) and (\ref{eq:scale00}),
we have the scaling
\begin{equation}
\Re\sigma_{xx,\phi}(\omega )=\frac{\rm const.}{\omega ^{\gamma}},
\quad {\rm for} \,\, \omega \to 0
\label{eq:unusual0}
\end{equation}
where
\begin{equation}
\gamma =\frac{\delta}{\eta}\simeq 0.55.
\label{eq:unusual1}
\end{equation}

The naive argument would lead to
$\Re\sigma_{xx,\phi}(\omega )\simeq c\,\delta(\omega)+c'/\omega$
when $E_F$ is in the energy band (metal) and
$\Re\sigma_{xx,\phi}(\omega ) \simeq c'/\omega$
when $E_F$ is in the energy gap (insulator) as $\omega$ tends to
zero.
Our result Eq.~(\ref{eq:unusual0}) is different from
both of these simple expectations. We shall give the correct analysis
below.
Let us assume that the matrix elements in Eqs.~(\ref{eq:Drude})
and (\ref{eq:scatter})
to be constant
$|v_{x}^{ll'}(\mbox{\boldmath $k$})|=v_x$.
Then the Drude weight $D(q)$ can be estimated as
\begin{eqnarray}
    D(q)\simeq \frac{1}{2} {v_x}^2
      \oint_{E^{m}(k_x,k_y)=E_F}dk
\frac{1}{|\nabla  E^{m}({\mbox{\boldmath $k$}})|}
        =2{\pi}^2 N^m (E_F){v_x}^{2}.
\end{eqnarray}
where
$N^m (E) =\frac{1}{(2\pi )^2}\oint_{E^{m}(k_x,k_y)=E}dk\frac{1}
{|\nabla  E^{m}|}$ is a density of states of the $m$-th band.
Since there are $q$ bands in the total spectrum,
integrated density of states for
each band is $1/q$ and $N(E_F) \sim \frac{1}{q\Delta E}$, where
$\Delta E$, the width of the band including
$E=0$, is of the order
of $\Delta E \sim q^{-\eta}$ (see Eq.~(\ref{eq:del})).
Also we may estimate $v_x =\Delta E/\Delta k_x$ where
$\Delta k_x \sim 1/q$ is a size of the magnetic Brillouin zone
in the $x$ direction. Thus we have
\begin{equation}
D(q) \simeq {\rm const.}\times q^{1-\eta}.
\end{equation}
It suggests a scaling relation
\begin{equation}
\alpha =\eta -1,
\end{equation}
which is consistent with the numerical result of $\alpha =0.828\pm0.002$
and $\eta =1.8285\pm 0.0006$. It gives the scaling relation between
the Drude weight and the spectrum. Note the scaling index for
the Drude weight is $\alpha$ and the scaling index for the spectrum is
$\eta$.
{}From Eq.~(\ref{eq:Drude00}) it implies
\begin{equation}
\delta_D=1.
\end{equation}

Next let us consider the contribution from the interband scattering
Eq.~(\ref{eq:scatter}).
Since the onset of $\Re\sigma^S_{xx,\phi_l}(\omega)$ is of
the order of $\Delta E$, we have
\begin{equation}
  \Re\sigma_{\rm ave}^S(\phi_l) \simeq \frac{1}{\Delta E}{v_{x}}^{2}N_F N_c,
\end{equation}
where $N_c (\sim O(1))$ is a number of a possible combination
of the energy bands which can contribute the process.
Thus we can estimate $\Re\sigma_{\rm ave}^S(\phi_l)$
as $\Re\sigma_{\rm ave}^S(\phi_l) \simeq \frac{1}{\Delta E}(q\Delta E)^{2}
\frac{1}{q\Delta E}=q$.
It suggests
\begin{equation}
\delta_S=1.
\end{equation}
\section{summary}
\label{sec:sum}
We have considered both of the longitudinal and
transverse conductivities
$\sigma_{\mu \nu,\phi}(\omega )$
of the two-dimensional
electrons on the square lattice at an arbitrary filling
factor. Especially the
longitudinal conductivity
$\sigma_{xx,\phi}(\omega )$ is discussed in details.
It consists of two terms:
the Drude term $\sigma_{xx,\phi}^D(\omega )$
and the interband scattering term $\sigma_{xx,\phi}^S(\omega )$.

We numerically calculated the real parts
$\Re\sigma_{xx,\phi_l}^D(\omega )$ and
$\Re\sigma_{xx,\phi_l}^S(\omega )$ for a series of
systems with rational fluxes $\{\phi_{l}\}$
which converge to the irrational value $\tau =\frac{\sqrt{5}-1}{2}$.
The case when the Fermi energy $E_F$ lies near
$E=0$ was treated in details.
Two parts of the longitudinal conductivities
$\Re\sigma_{xx,\phi_l}^D(\omega )$ and
$\Re\sigma_{xx,\phi_l}^S(\omega )$
show
the same scaling behavior in the incommensurate limit.
Taking into account
the self-similar structure of the energy spectrum,
we performed the averaging procedure near $\omega =0$.
It revealed the clear scaling behavior
$\Re\sigma_{xx}(\omega) \sim {1}/{\omega ^{\gamma}}
\,(\gamma \simeq 0.55,\omega \to 0)$.
This result is quite distinct from the commensurate case.
%
%


%

%
%

%
%
%
\begin{figure}
\caption{Drude weights $D(q)$ for systems with
three Fermi energies $E_F$ in the central band;
(a) $E_F=0$, $\phi=\frac{3}{5}$, $\frac{13}{21}$,
$\frac{55}{89}$, and $\frac{233}{377}$,
(b) $E_F=0$, $\phi=\frac{8}{13}$, $\frac{34}{55}$,
$\frac{144}{233}$, and $\frac{610}{987}$,
(c) $E_F=-0.475\Delta_l$, $\phi=\frac{3}{5}$,
$\frac{13}{21}$, $\frac{55}{89}$, and $\frac{233}{377}$,
(d) $E_F=-0.475\Delta_l$, $\phi=\frac{8}{13}$,
$\frac{34}{55}$, $\frac{144}{233}$, and $\frac{610}{987}$,
(e) $E_F=-0.4995\Delta_l$, $\phi=\frac{3}{5}$,
$\frac{13}{21}$, $\frac{55}{89}$, and $\frac{233}{377}$, and
(f) $E_F=-0.4995\Delta_l$, $\phi=\frac{8}{13}$,
$\frac{34}{55}$, $\frac{144}{233}$, and $\frac{610}{987}$
where $\Delta_l$ is a width
of the central band.}
\end{figure}

\begin{figure}
\caption{$\Re\sigma_{xx,\phi_l}^S(\omega )$ for systems with
rational fluxes $\{\phi_l\}$ with $E_F=0$;
(a) $\phi_{5} =8/13$, (b) $\phi_{6} =13/21$, (c) $\phi_{8} =34/55$,
(d) $\phi_{9} =55/89$, (e) $\phi_{11} =144/233$
and (f) $\phi_{12} =233/377$.
Note the differences in the energy scales.}
\end{figure}

\begin{figure}
\caption{$\Re\sigma_{xx,\phi_l}^S(\omega )$ for systems with
rational fluxes $\{\phi_l\}$;
(a) $\phi_{5} =8/13$, (b) $\phi_{6} =13/21$, (c) $\phi_{7} =21/34$,
(d) $\phi_{8} =34/55$,
(e) $\phi_{9} =55/89$, (f) $\phi_{10} =89/144$, (g) $\phi_{11} =144/233$
and (h) $\phi_{12} =233/377$.
The Fermi energy $E_F$ for each of them lies in the gap just above $E=0$.
Note the differences in the energy scales.}
\end{figure}

\begin{figure}
\caption{Averaged conductivities  $\Re\sigma_{\rm ave}^S(\phi_l)$
in two cases; (a)
when $E_F$ is in the energy gap,
and (b) when $E_F$ is in the energy band.
$\Re\sigma_{\rm ave}^S(\phi_l)$ was calculated
for three values of $W_0$
which give
$M=3, 7,$ and $11$, respectively;
(a)
(1)$M=3$, $\phi=\frac{8}{13}$, $\frac{34}{55}$, $\frac{144}{233}$,
(2)$M=3$, $\phi=\frac{13}{21}$, $\frac{55}{89}$, $\frac{233}{377}$,
(3)$M=3$, $\phi=\frac{21}{34}$, $\frac{89}{144}$,
(4)$M=7$, $\phi=\frac{8}{13}$, $\frac{34}{55}$, $\frac{144}{233}$,
(5)$M=7$, $\phi=\frac{13}{21}$, $\frac{55}{89}$, $\frac{233}{377}$,
(6)$M=7$, $\phi=\frac{21}{34}$, $\frac{89}{144}$,
(7)$M=11$, $\phi=\frac{8}{13}$, $\frac{34}{55}$, $\frac{144}{233}$,
(8)$M=11$, $\phi=\frac{13}{21}$, $\frac{55}{89}$, $\frac{233}{377}$, and
(9)$M=11$, $\phi=\frac{21}{34}$, $\frac{89}{144}$,
and (b)
(1)$M=3$, $\phi=\frac{8}{13}$, $\frac{34}{55}$, $\frac{144}{233}$,
(2)$M=3$, $\phi=\frac{13}{21}$, $\frac{55}{89}$, $\frac{233}{377}$,
(3)$M=7$, $\phi=\frac{8}{13}$, $\frac{34}{55}$, $\frac{144}{233}$,
(4)$M=7$, $\phi=\frac{13}{21}$, $\frac{55}{89}$, $\frac{233}{377}$,
(5)$M=11$, $\phi=\frac{8}{13}$, $\frac{34}{55}$, $\frac{144}{233}$, and
(6)$M=11$, $\phi=\frac{13}{21}$, $\frac{55}{89}$, $\frac{233}{377}$.}
\end{figure}
\end{document}